# An array of microresonators as a photonic extreme learning machine

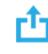

View Online   Export Citation   CrossMark

Stefano Biasi,[a])  Riccardo Franchi,  Lorenzo Cerini,  and Lorenzo Pavesi

## AFFILIATIONS

Nanoscience Laboratory, Department of Physics, University of Trento, 38123 Trento, Italy

[a])Author to whom correspondence should be addressed: stefano.biasi@unitn.it

## ABSTRACT

Machine learning technologies have found fertile ground in optics due to their promising features based on speed and parallelism. Feed-forward neural networks are one of the most widely used machine learning algorithms due to their simplicity and universal approximation capability. However, the typical training procedure, where all weights are optimized, can be time and energy consuming. An alternative approach is the Extreme Learning Machine, a feed-forward neural network in which only the output weights are trained, while the internal connections are random. Here we present an experimental implementation of a photonic extreme learning machine (PELM) in an integrated silicon chip. The PELM is based on the processing of the image of the scattered light by an array of 18 gratings coupled to microresonators. Light propagation in the microresonator array is a linear process while light detection by the video camera is a nonlinear process. Training is done offline by analyzing the recorded scattered light image with a linear classifier. We provide a proof-of-concept demonstration of the PELM by solving both binary and analog tasks, and show how the performance depends on the number of microresonators used in the readout procedure.

## I. INTRODUCTION

In the last decade, the implementation of machine learning methods in photonic systems has rejuvenated optical computing.[1] Photonics represents a suitable platform for artificial intelligence due to its operational speed, low power dissipation, and versatility of optics, ranging from the simple complex nature of the optical field[2] to the possibility of using multiplexing techniques.[3] Photonic neural networks (PNN) have been demonstrated to operate with state-of-the-art performance in challenging tasks such as image classification[4–6] and non-linear decision boundary.[2] Among PNNs, photonic integrated circuits (PIC) are particularly attractive because they can meet the demand for scalability of neural networks by mitigating problems related to control, power consumption, and physical size.[7–9] Feed-forward neural networks (FFNN) based on cascades of Mach-Zehnder interferometers (MZIs) have been demonstrated on-chip,[10] showing excellent performance in vowel recognition. However, the training process is time and power consuming as FFNNs generally require optimization of the response through slow gradient descent algorithms. A promising alternative to these approaches is PNNs that do not require full control of the network, such as Reservoir Computing (RC)[13–15] and Extreme

Learning Machines (ELMs).[16,17] ELMs are FFNNs composed of a single hidden layer in which training occurs only at the readout. The underlying operation is based on a nonlinear mapping of input information to a higher dimensional space, on which a linear classifier defines the readout layer in the training and testing phase. The potential of ELM has been demonstrated in bulk systems using different information mapping strategies such as coherent light phase modulation with a spatial light modulator,[18] frequency multiplexing by a phase modulator acting on a monochromatic source,[19] speckle patterns resulting from multiple scattering[20,21] and time-multiplexed fiber loops.[22] PIC implementation of ELM is largely unexplored, while few PICs where only a subset of connections are trained have been studied.[23–27] This is probably due to the fact that integrated implementations of ELM lead to a limited increase in the spatial dimensions of the mapping with respect to a bulk approach.

In this work, we propose and experimentally validate an implementation of the ELM in a silicon integrated photonic circuit: a photonic ELM (PELM). The PELM is based on an array of 18 microresonators arranged in a geometry that forces a single propagation direction for the optical signal. Both the encoding of information and its processing take place in the PIC, while the training and testing are performed offline. The PELM is tested with two-level and multi-level tasks. Therefore, we show the performance of the PELM in solving linear and nonlinear logical operations and in two classification problems: the recognition of iris flowers and the banknote authentication. We also demonstrate how network performance depends on the number of physical nodes used in the readout during the training and testing.

## II. EXPERIMENTAL IMPLEMENTATION

### A. Optical circuit

An ELM is a FFNN in which the input information is processed and sent through a single hidden layer to at least one output node.[16] The hidden layer forms the computational reservoir and does not require training of its response. In contrast, training takes place in the output layer. As shown in Fig. 1(a), a given dataset of $N$ features, $\mathbf{X}$, is injected into the hidden layer. Here, the input information is mapped to a higher dimensional space. This space is obtained by linear combinations of the input data with unknown random weights. An infinitely differentiable nonlinear activation function projects the data into an $\mathbf{H}$ matrix that represents the output of the hidden layer. Then, the output of the ELM, $\mathbf{Y}$, is formalized as the linear combination of $\mathbf{H}$ with a vector of weights $\boldsymbol{\beta}$, i.e., $\mathbf{Y} = \mathbf{H}\boldsymbol{\beta}$. Training is performed only on the output response by determining the weights $\boldsymbol{\beta}$ that allow the reproducing a target, $\mathbf{T}$. The vector of optimal output weights is the least squares solution of $\boldsymbol{\beta} = \mathbf{H}^T\mathbf{T}$, where $\mathbf{H}^T$ is the generalized Moore–Penrose inverse of $\mathbf{H}$. Note that ELM has the universal approximation ability[16] and, therefore, can reproduce any output with enough nodes in the hidden layer.

The PIC in Fig. 1(b) implements the PELM. The red lines represent the $450 \times 220$ nm$^2$ silicon waveguides embedded in silica cladding, while the gray triangles represent the grating couplers. The PIC was fabricated at the IMEC/Europractice facility within a multi-project wafer run and it is placed within a ceramic electronic packaging. The PELM has an input layer, where information is encoded in the optical field amplitude, and a hidden layer formed by an array of 18 microresonators coupled to output gratings. At the input, the signal is coupled through a grating coupler and split into four input waveguides by a balanced $1 \times 4$ multi-mode interferometer. Each of the four signals passes through a balanced MZI, where one of its two arms has a phase shifter (PS) actuated by a microheater. This consists of a straight strip of titanium nitride (TiN) with a length of about 60 $\mu$m and a width of about 6 $\mu$m, represented by the brown lines in Fig. 1(b). The current applied to the microheater induces a change in the refractive index, which in turn affects the interference at the output of the MZI. In this way, the input data are encoded into a four dimensional feature space. The four modulated input optical signals are denoted in Fig. 1(b) by the Roman numerals I, II, III, and IV. They excite a 3 × 3 array of coupled microresonator pairs, which form the hidden layer. Neglecting the backscattering due to surface-wall roughness, the coupled microresonator pair allows realizing the feed forward scheme since propagation forward direction of the input optical signal is maintained across the array. The microresonators have a vertically elongated racetrack geometry obtained by joining four Euler curves (minimum radius 15 $\mu m$ and angle 90°) to two pairs of straight sections of 4 $\mu m$ length. Euler curves were used to further reduce the backscattering phenomena. The microresonators have a quality factor of about $10^4$. Each of them has a tap located in the straight section that collects the 5‰ of the circulating intensity by scattering it off the chip through a grating coupler. A camera record the scattered light intensity image emerging from the microresonator array. In this way, the array of microresonators coupled to the gratings linearly maps the four input optical signals into 18 output scattered light images. An image processing step allows us to obtain **H**, which is composed of the values obtained by digitizing the intensity image of the scattered light. The hidden layer of the PELM thus consists of the linear optical propagation in the array of 18 microresonators and of the nonlinear detection (light intensity is the square of the optical field) of the scattered light by the 18 gratings. For monitoring and calibration purposes, each of the four input waveguides ends at a grating coupler, which allows the collecting of the array transmission optical signals. The output of the PELM is obtained in the readout layer via digital hardware, which applies the vector of weights **β** to the digitized intensities. The training is performed by the ridge regression, which finds the vector **β** that minimizes the regularized least square error $\|\mathbf{H}\boldsymbol{\beta} + \beta_0\mathbf{I} - \mathbf{T}\|^2 + \lambda^2\|\boldsymbol{\beta}\|^2$, where $\lambda$ is the regularization parameter defined by a five-fold cross-validation (for details on the software implementation of ridge regression, see Appendix A).

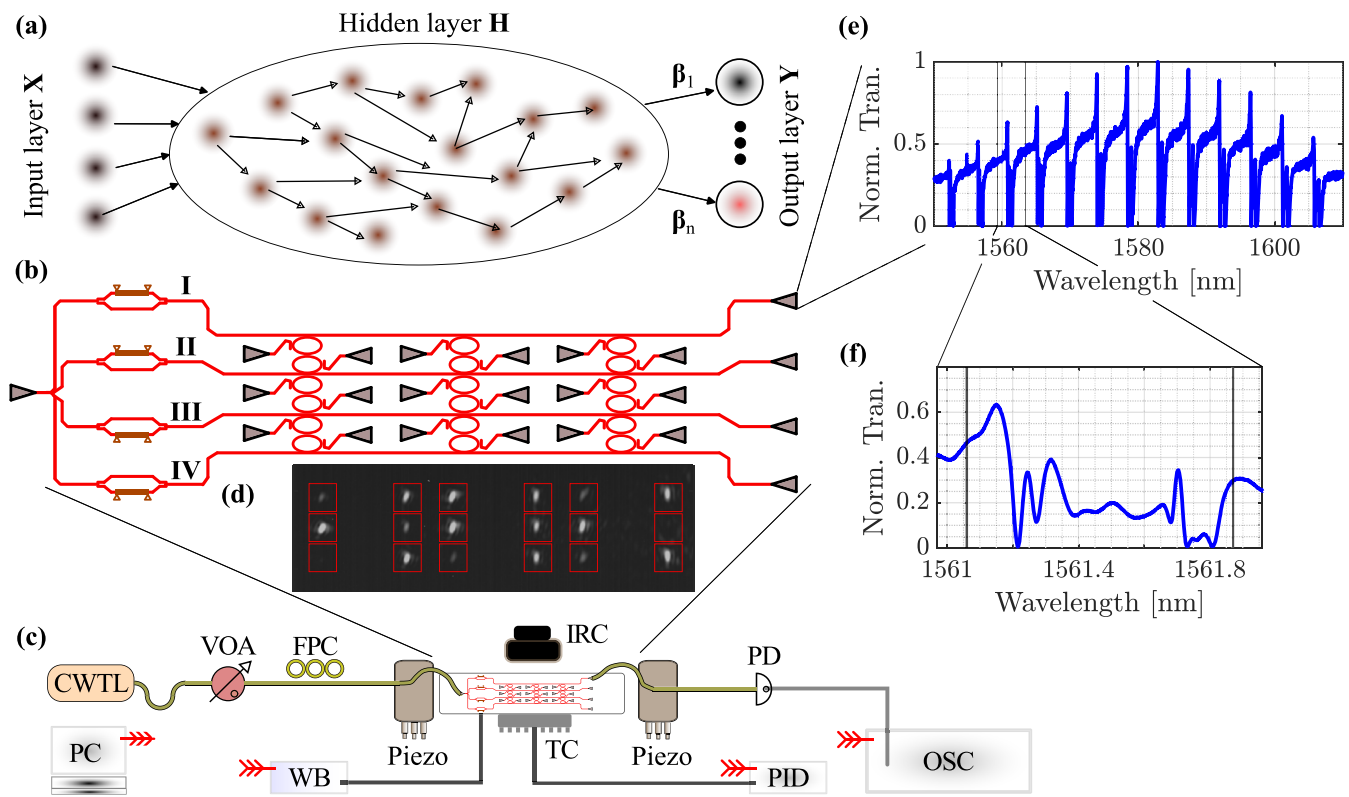

FIG. 1. (a) Sketch of the architecture of an Extreme Learning Machine (ELM), where an input dataset $\mathbf{X}$ is sent to a reservoir that generates the $\mathbf{H}$ matrix of the hidden layer. The output $\mathbf{Y}$ is the linear combination of $\mathbf{H}$ and the trainable vector of weights $\boldsymbol{\beta}$. (b) Design of the integrated neural network to implement the photonic ELM, the PELM. It consists of an input layer where the information encoding takes place, a hidden layer composed of an array of microresonators coupled to grating couplers, and a set of end grating couplers to calibrate the response. (c) Experimental setup. The different symbols are defined in the text. (d) Image captured by the infrared camera. The red squares highlight the light scattered by the grating couplers connected to the microresonators. (e) Normalized transmittance as a function of input wavelength. The light is coupled to the input grating and measured on the first calibration grating highlighted by the black dashed lines. (f) Zoom on the spectral response around 1561 nm. The two black vertical lines define the "in-resonance" and "out-of-resonance" operating wavelength ranges.

A key aspect of the ELM is the random nature of the weights applied to the linear combination that maps the input space to a higher dimensional output space. In our PIC, the mapping is done by coupling the four input waveguides to the microresonator array. Both the wavelength resonances (i.e., the diameters) of the microresonators, the couplings between microresonator and bus waveguide (M-W), the couplings between microresonator and microresonator (M-M), and the couplings between bus waveguide, microresonator and microresonator (M-W-M) determine the intensities of the optical signals that propagate in the microresonator array. Expressed in other terms, the weights of the linear combination of the input information into the hidden layer. In the PIC design, the size of the microresonators and the gaps are all nominally equal, specifically, the values used in the design are shown in Table II of Appendix B. Therefore, in our system, the randomness is only given by the stochastic fabrication errors.

### B. Experimental setup and H-matrix generation

The PELM has been tested using the experimental setup shown in Fig. 1(c). The laser source is a fiber-coupled continuous-wave tunable laser (CWTL: Yenista OPTICS TUNICS-T100) operating in the infrared range (1470–1580 nm). Its emission passes through a variable optical attenuator (VOA) and a fiber polarization control stage (FPC). The generated continuous-wave signal is coupled to the PIC via a single mode stripped fiber. Another single mode stripped fiber collects the output from the calibration grating couplers. The collected signal is finally sent to an InGaAs photodetector (PD: Thorlabs, PDA20CS2), whose response is monitored by an oscilloscope (OSC: PicoScope 4000 series). Both the input and output stripped fibers are placed on a three-axis linear piezoelectric stage for proper alignment. A write board (WB: Measurement Computing USB-3106) applies currents to the microheaters of the input layer and controls the voltage that regulates the VOA. The temperature of the PIC is kept constant by a thermostat holder controlled by a proportional-integral-derivative (PID: SIM960 Analog controller) connected to a Peltier cell and a 10 k$\Omega$ thermistor. This allows the working point of the microresonator array to be controlled, ensuring excellent reproducibility of measurements by mitigating the impact of the environment.

The scattered light from the 18 grating couplers coupled to the microresonators is monitored by an infrared camera (IRC: Xenics XEVA-CL-640-25) mounted on a continuous zoom objective (Navitar). A three-axis linear stage ensures that the objective is placed over the PIC and maintains its position during the PELM training and testing. The grayscale image captured by the camera has a resolution of 512 × 640 pixels with 16 bit per pixel. An example image is shown in Fig. 1(d). Here, the red squares mark the positions of the 18 grating couplers, and thus the encircled bright spots represent the intensities of the optical signals propagating in the different microresonators. The readout and generation of **H** are done by digitally processing the captured images. This procedure can be performed by considering only the 18 physical nodes and or by exploiting virtual nodes. In the first case, the value attributed to each node is calculated by averaging the 1296 pixels contained in each red square. In the second case, the number of nodes is virtually increased by performing different operations on the image. For example, each individual pixels can be considered as a node or additional functions are applied to the 1296 pixels contained in the square, such as the standard deviation, the median, or the geometric mean. Note that the objective of the camera is set to image only the array of microresonators and, therefore, a single pixel corresponds to an area of about 1.5 × 0.5 $\mu m^2$ on the PIC.

Encoding the signal in the PIC by varying the current of the PSs with the write board can introduce thermal cross-talk. Such an effect on the dynamics of microresonators has recently been studied in our work.[28] It was shown that the variation of the microheater induces a global heat flow through the substrate, which delays the reaching of the equilibrium state of the system. In the case of a packaged PIC (ceramic package), a microresonator is influenced by a microheater distant more than 200 $\mu m$ away with a relaxation time of about 220 ms and reaches the steady state in about 1 s. Since in the PELM the distance between the microresonator and the nearest microheater is about 250 $\mu m$, the overall thermal effect is not negligible. On the other hand, the camera cannot acquire and store images at a rate higher than 100 ms/frame. Therefore, it is necessary to impose a guard time between the change of the input and the image acquisition. To speed up image acquisition, two images are acquired at 300 ms intervals for each new input (i.e., different current

14 September 2023 07:13:43

injection in the microheaters). If they differ, an additional 300 ms pause is added. This process is repeated until the two acquired images are similar. The definition of similarity is given by a threshold of percentage difference in the pixels of the two images. This percentage value varies according to the problem addressed, depending on factors such as the integration time of the camera and the database to be encoded. As a result, its estimation requires an experimental test measurement prior to the task execution.

## III. RESULTS

### A. Setting the working point

The performance of the network depends on the wavelength of the optical input signal. The encoding of the input information, determines the distribution of the signal in the array of microresonators. Figure 1(e) shows the normalized transmittance as a function of the incident wavelength for an incident laser power of 1 mW. The optical signal is coupled to the input grating and the transmittance is monitored in the first calibration grating, indicated by the dashed black lines in Fig. 1. Scanning is performed at 5 nm/s by means of the Picoscope, which allows synchronization of the laser wavelength with the data acquired by the photodetector. The transmittance shows a broad bell pattern, typical of grating couplers (coupling losses of about 6 dB, equally divided between input and output gratings). The spectrum is characterized by a series of bands, each one consisting of a series of resonant dips. Figure 1(f) shows a zoom of the response around 1561 nm resonance. The presence of fabrication errors induces a variation of the geometrical parameters of the microresonator array, which yields several local minima in the resonance band.[29] This demonstrates the needed randomness of the microresonator array. In the following tasks, we have used optical signal wavelengths about the wavelength interval 1561.06–1561.87 nm, which corresponds to a resonant band. We use the term "in-resonance" for wavelengths included in this interval and "out-of-resonance" for the others. In the following, the incident laser power is set to 1 mW. This value ensures a linear response of the microresonators array.

### B. Binary input

The first tasks we addressed are the two-level logical operations AND, OR, and XOR, for which the truth tables are given in Table I. AND and OR are linearly separable tasks and, therefore, do not require a nonlinear activation function to be solved. In contrast, XOR is a nonlinear task. We do not use the logical operations as a network benchmark, but rather as a proof-of-concept demonstration of the use of the PELM architecture.

The encoding is implemented by using only two of the four MZIs of the input layer. The remaining pair is set to have destructive interference and, thus, a null signal at the output. As a result, the task is performed using only two excitation waveguides of the microresonator array, using the maximum of the signal amplitude as bit 1 and 30% of the maximum as bit 0. The network is tested by implementing all six possible permutations of the input signals (I, II, III, and IV) in the encoding phase. Specifically, each input signal pair reproduces four times the excitation combination formed by the four bit pairs $\{(0,0),(0,1),(1,0),(1,1)\}$. For each combination, the bit pairs are randomly generated from the input layer. Only the **H** matrices obtained from a single excitation combination (four pairs) are used to train the PELM, while the remaining measurements are used as tests, resulting in a training/test ratio of 25/75. The generation of the **H** matrix is done by using the 18 physical nodes. The input set is randomly replicated several times, both to test the reproducibility of the neural network and to account for small fluctuations in node intensities due to noise in the training process.

**TABLE I.** Truth table for AND, OR, XOR operations.

| Inputs | AND | OR | XOR |
|--------|-----|-----|-----|
| (0,0) | 0 | 0 | 0 |
| (0,1) | 0 | 1 | 1 |
| (1,0) | 0 | 1 | 1 |
| (1,1) | 1 | 1 | 0 |

Logical operations are undemanding tasks for the PELM whose accuracy is 100% for many in-resonance and out-of-resonance wavelengths. Consequently, the performance of the PELM is evaluated in terms of the distance between output values that should be 1 and those that should be 0. Whenever a threshold can be drawn between these two outcomes, the network solves the task. We define the smallest distance between these two results as the gap size,

$$Gs[\lambda] := \frac{\min_{T=1}[\mathbf{Y}[\lambda]] - \max_{T=0}[\mathbf{Y}[\lambda]]}{\max[\{\max[\mathbf{Y}[\lambda]] - \min[\mathbf{Y}[\lambda]]\}, 1]}. \quad (1)$$

$Gs[\lambda]$ gives the normalized value of the difference between the minimum of the responses that should be 1 and the maximum of the responses that should be 0. As a result, $Gs$ greater than 0 means that the network solves the task. Conversely, $Gs$ smaller or equal than 0 means that the network does not solve the task for at least one pair of excitation combinations. The greater the value of $Gs$, the better the performance of the PELM and the robustness of the accuracy against noise. Note that in Eq. (1), the scaling factor is defined by the maximum of the set composed by the difference between the maximum and minimum response and 1.

The top graphs in Fig. 2 show $Gs[\lambda]$ for the AND, OR, and XOR in the best input signal pair configuration. For AND, OR, and XOR, the best input pairs are (II, IV), (II, III), and (III, IV), respectively. In the plots, the vertical black lines highlight the resonance region of the microresonator array shown in Fig. 1(f). The network is tested with 76 values of incident wavelengths distributed in-resonance and out-of-resonance. The horizontal red lines highlight the zero value of $Gs$. In all three logical operations, the PELM fails to solve the task for at least one pair of the excitation combinations at just one out-of-resonance wavelength. Thus, for in-resonance wavelengths, the logical operations are always solved by the PELM. For the AND and OR, the best performance is reached in-resonance at 1561.68 and 1561.22 nm, respectively. While for the XOR the best performance is near the upper limit of the resonant region, which is at 1561.9 nm. The bottom graphs in Fig. 2 show the test result for the in-resonance wavelength that has the maximum $Gs$ value. Specifically, the plots display the PELM output for all the 16 random bit pairs contained in the four repetitions of the excitation combination. The red dots are the outputs that should be 1, while the blue dots are those that should be 0. The PELM reproduces the targets perfectly by showing a clear distinction between bits 0 and 1.

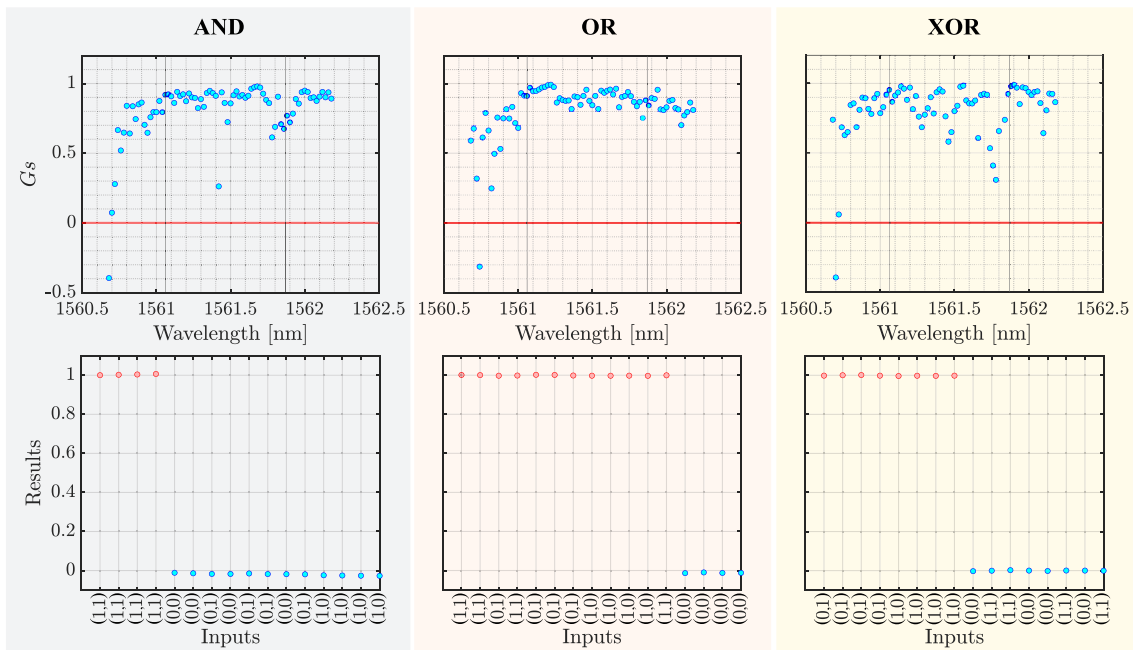

**FIG. 2.** Top, gap size ($Gs$) as a function of incident wavelengths for the AND, OR, and XOR operations. For each operation, the experimental data are obtained from the best input signal configuration. In the plots, the black lines mark the in-resonance and out-of-resonance regions, while the red horizontal lines mark $Gs = 0$. The network solves these binary tasks when $Gs > 0$. Bottom, outputs of the neural network in the test procedure for the in-resonance wavelength that has the best $Gs$ value. The inputs are all the 16-bit pairs of the four repetitions of the excitation combination. The red and blue dots are the bits that should be one and zero, respectively.

The performance of an ELM is strongly dependent on the number of nodes in the hidden layer. Previous results show that 18 nodes allow the PELM to solve the logical operations both in-resonance and out-of-resonance. However, the lower the number of physical and/or virtual nodes, the higher the operating speed of the network, albeit at the expense of its performance. Thus, depending on the tasks to be solved, there is an optimal value of nodes that maximizes performance while minimizing training time. The digital hardware readout system allows changing the number of nodes used to generate **H**. We studied the dependence of the PELM's performance on the number of nodes for the XOR operation in the input pair configuration (III, IV) and considering only the in-resonance wavelength (bottom right panel in Fig. 2).

Formally, we computed $Gs$ for all $n$-combination of the 18 physical nodes, where $n = 1, 2, 3, \ldots, 18$. An $n$-combination is a subset of $n$ different elements of the 18 physical nodes. Two $n$-combinations are identical if and only if they consist of the same elements. The number of $n$-combination is given by the binomial coefficient. This value is 18 and 1 for $n = 1$ and $n = 18$, respectively. While it reaches the maximum value of 48 620, for $n = 9$. Figure 3(a) shows the values of $Gs$ as a function of the physical nodes. For each node number $n$, the red squares and green downward triangles show the $Gs$ obtained from the $n$-combination that gives its maximum value and the one that gives its minimum value. In addition, the black dots show the average value of $Gs$ for each $n$, obtained by using the results of all $n$-combinations. In the best combination, the PELM solves the task with a single physical output node. In contrast, in the worst combination, the PELM shows a positive value of $Gs$ with at least eight nodes and shows near maximum performance with 16 nodes. Looking at the mean value of $Gs$, the PELM solves the XOR with a number of nodes equal to 3. The curve rises rapidly to five physical nodes and then tends more slowly to a plateau characterized by the maximum value of 1. Consequently, 18 nodes are excessive for the XOR. In fact, in the best $n$-combination with $n = 3$, the network solves the XOR with a $Gs$ comparable to that obtained with all 18 nodes.

Interestingly, there are configurations of input signals and in-resonance wavelengths where the PELM solves the XOR with only one microresonator, even without performing the ridge regression in the readout. An example is shown in Fig. 3(b), for the input signal pair (II, IV) and an in-resonance wavelength of 1561.56 nm. The four images correspond to the four pairs (0, 0), (1, 1), (01), and (1, 0). Focusing on the grating coupler indicated by the yellow star, one observes that the intensity circulating in the coupled microresonator resolves the XOR. In fact, the (0, 1) and (1, 0) pairs show a clear light spot. In contrast, for the (0, 0) and (1, 1) pairs, the microresonator is unloaded (destructive interference) and the light emitted by the grating is zero. This behavior matches the logical XOR operation (Table I).

## C. Analog input

The analog input tasks we addressed are two well-known toy datasets in machine learning: iris flower classification[30] and banknote classification.[31] We do not intend to use them as a benchmark against other neural network systems, but as a testbed

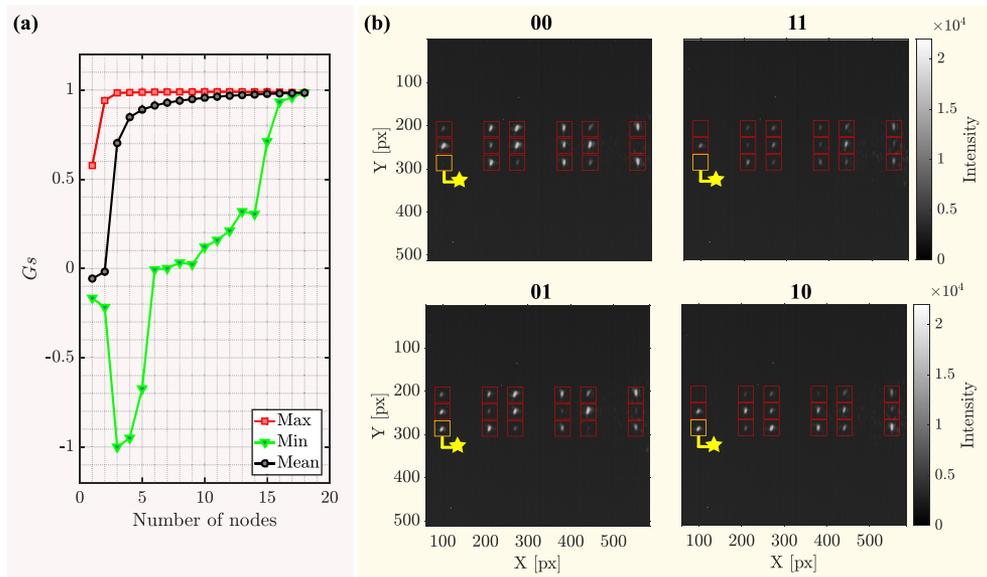

**FIG. 3.** (a) *Gs* as a function of the number of the used output nodes *n* for the XOR operation. The red squares and the green downward triangles show the *Gs* for the *n*-combination, which gives its maximum and its minimum value, respectively. The black dots denote the average of *Gs* obtained by considering for each *n* value all the *n*-combination. (b) Images acquired by the infrared camera for the pairs of the excitation combination, namely (0, 0), (1, 1), (0, 1), and (1, 0). This result is obtained for an in-resonance wavelength by using the input signal pair (II, IV). The yellow stars indicate the grating couplers, which mimic the outcomes of the XOR operation without performing the ridge regression in the readout layer.

for our architecture on inputs encoded with four multi-level features.

The *iris flower* database consists of 150 flowers described by four input features characterized by real values. Physically, the four features correspond to the length and width of the petals and sepals of the flower. The goal is to classify a given flower into one of three possible subspecies called: Setosa, Versicolor, and Verginica. These three classes are equally distributed among the 150 flowers in the database. The classification is not trivial, because even with all four inputs, only the Setosa species can be separated linearly from the others.

The iris flowers are sent to the hidden layer using all four MZIs of the input layer. As a result, each feature is encoded on a given input signal [indicated in Roman numerals in Fig. 1(b)], and its values are scaled between the maximum and 30% of the maximum signal amplitude. Sepal length, sepal width, petal length, and petal width have 35, 23, 43, and 22 different values, respectively, resulting in as many levels of signal amplitude in our encoding. Each flower in the database is encoded and sent to the hidden layer ten times in a random order. This procedure is performed for 22 wavelengths of the incident laser signal, distributed over 11 in-resonance and 11 out-of-resonance values.

For each wavelength, the performance of the network is studied. From the 1500 acquired images, the corresponding **H** matrices are extrapolated, consisting only of the 18 physical nodes. They are distributed into a training set and a test set with a percentage of 70% and 30%, respectively. This distribution is performed using the Monte Carlo cross-validation[52] to avoid problems such as selection bias or overfitting. Specifically, for each wavelength, the test and training are performed 100 times by resampling the test and training set differently. The different categories can be assigned in the readout layer through two procedures: training a linear classifier for each species (three basis) or training a single linear classifier for all three species (single basis). In the first procedure, the final decision is made according to a winner takes all scheme.[33] Consequently, the classifier that provides the greatest value is the one that defines the species of the flower under test. In the second procedure, the single classifier provides a specific value for each species. As a result, the PELM has a single output for each flower. This output directly defines the species. Note that the readout processing was performed on both the hidden layer output and the raw input data.

Figures 4(a) and 4(b) show the classification rate as a function of the incident wavelength by training the network with three basis and with one single basis, respectively. The black and blue dots show the mean values over the 100 repetitions, with the corresponding variance indicated by the error bars. The horizontal rectangles show the mean and variance of the classification rate over 100 repetitions performed on the raw data, i.e., the input. Thus, this value represents our theoretical limit obtained by simply applying the linearization to the input database.

Using the first method [three basis, see Fig. 4(a)], the PELM improves the classification rate by more than 7% over the theoretical limit for both in-resonance and out-of-resonance wavelengths. Specifically, the theoretical limit is $(83.6 \pm 0.5)\%$, while the maximum classification rate is $(96.8 \pm 0.2)\%$ obtained in-resonance at 1561.45 nm. This value is comparable to the result obtained by Lupo *et al.*[19] In this work, a bulk implementation of the ELM based on wavelength multiplexing is reported, which achieves a maximum classification rate of 97.7% (experiment) or 96.3% (simulation). Using the second method [one basis, see Fig. 4(b)], the PELM improves the classification rate with respect to the theoretical limit only for three in-resonance wavelengths. The theoretical limit is $(96.6 \pm 0.3)\%$ while the best PELM classification rate is $(97.2 \pm 0.2)\%$ at 1561.45 nm. Note that this in-resonance wavelength is the same wavelength as the maximum classification rate achieved with the three basis.

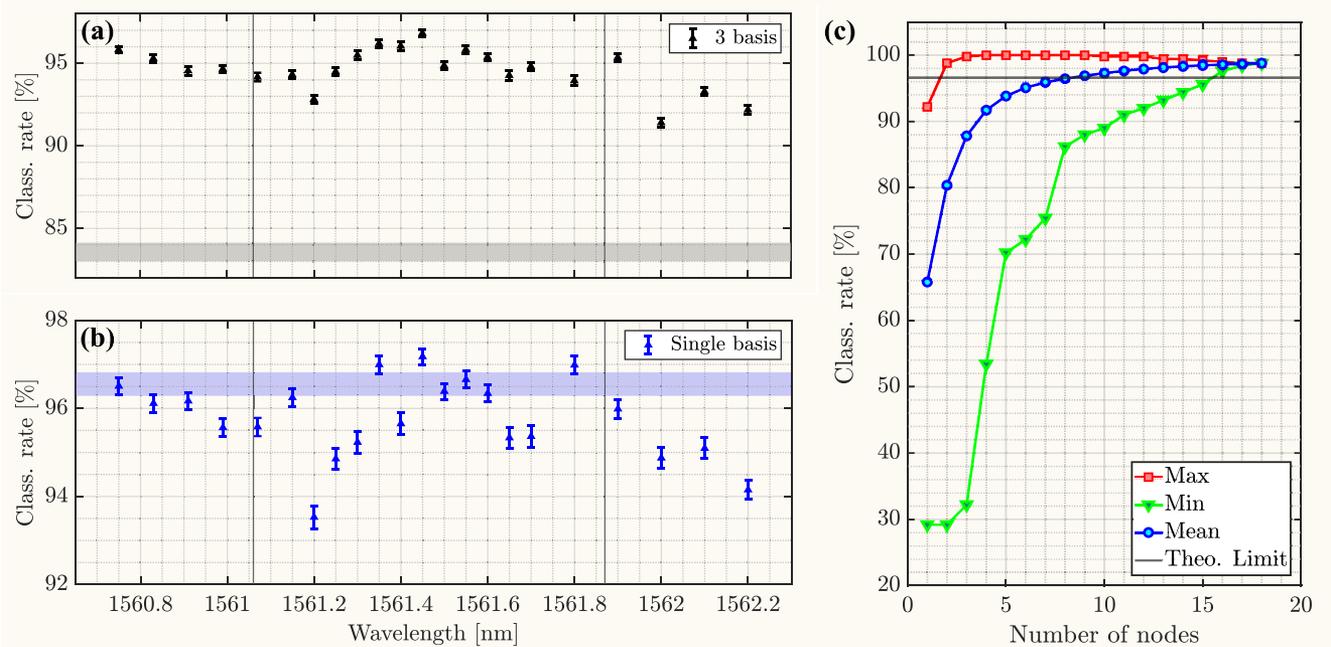

**FIG. 4.** Classification rate as a function of input wavelength, computed by using three linear classifiers (a) and one linear classifier (b) in the readout layer. The gray and the light blue rectangles show the theoretical limit obtained by applying the readout procedure to the raw database. The maximum classification rate is $(96.5 \pm 0.2)\%$ and $(97.2 \pm 0.2)\%$ for (a) and (b), respectively, reached at 1561.45 nm. (c) Classification rate as a function of physical node number. The red squares and downward triangles show the results of the $n$-combination that induces the best and worst classification rate. For each node number, the blue dots indicate the average classification rate overall $n$-combination. The theoretical limit is highlighted by the horizontal gray line.

It is worth noting that the single classifier [see Fig. 4(b)] provides a classification rate on the raw data comparable to the maximum obtained by the PELM in-resonance by using the three basis with the winner takes all scheme [see Fig. 4(a)]. This significant performance difference can be seen by comparing the two theoretical limits. It depends strongly on the database to be tested. This is shown in Appendix A, in which we study a database consisting of three three-dimensional spheres associated with three different colors. Although the spheres are linearly separable, for some configurations the three basis allows the spheres to be classified while the single does not, and for other configurations the reverse is true. This is a limitation of the ridge regression algorithm that can be avoided by using multinomial logistic regression. Even though this result is well known in the field of machine learning,[34] the classification tasks by multi-basis via ridge regression is often found in the literature.[18,19,26,35] This may be related to the study of network performance improvement on raw data rather than the search for the best absolute classification rate. In addition, linear ridge-based classifiers are typically faster than a multinomial logistic regression, especially in the case of a single basis. The classification rate as a function of wavelength using the multinomial logistic regression as linear classifier for the Iris flower is reported in Appendix C.

Note that with a linear classifier, the classification result also depends strongly on the permutation of the target, i.e., on how the values identifying the three species are distributed in the target vector $\mathbf{T}$. In our discussion, we used one of the permutations that give the maximum value of the theoretical limit. Setting 1, 2, and 3 for Setosa, Versicolor, and Verginica, respectively, the best classification rate is given by using $\mathbf{T} = (1, 2, 3)$ or $\mathbf{T} = (3, 2, 1)$.

Following the same procedure as described for the XOR operation (Subsection III B), we studied the performance of the PELM as a function of the number of nodes. We calculated the classification rate for all $n$-combinations of the 18 physical nodes. In this study, the one basis method is applied by using 70% of the database for the training procedure. Due to computational reasons, Monte Carlo cross-validation for the data distribution between the test and training sets is not possible. In fact, a PC equipped with a 32-core CPU (E5-2687W) requires three days to provide the results of a single iteration. Thus, a Monte Carlo cross-validation with 100 repetitions would require almost a year of computing time. Figure 4(c) shows the classification rate as a function of the number of nodes. For each value of $n = 1, 2, 3, \ldots, 18$, the red squares and the downward green triangles refer to the classification rate computed from the best and the worst $n$-combination, respectively. The blue circles to the average classification rate. The horizontal gray line denotes the theoretical limit obtained by applying the linear classifier to the input.

Thus, points with a value above this line show an improvement in the classification rate due to the PELM. We see that the PELM average result exceeds the theoretical limit with nine physical nodes. In addition, the best configuration shows excellent performance with only two physical nodes, while at least 16 nodes are needed for the worst *n*-combination.

The *banknote authentication* database consists of 1372 banknotes described by four input features.[31] Physically, the four features are the values obtained by mathematical functions applied to the banknote images. For each banknote, they report the value of the entropy of the image, the variance, the skewness, and the kurtosis of the wavelet-transformed image. The purpose is to classify a given banknote as genuine or counterfeit.

The banknote task is implemented in the PELM similarly to the method used for the iris flower task. As a result, the banknotes are sent to the hidden layer by encoding each feature on a corresponding input signal. The values of each feature are scaled between the maximum and 30% of the maximum signal amplitude. However, the number of distinct values contained in each feature is significantly larger than in the Iris flowers case. In fact, the variance/skewness/kurtosis of the wavelet-transformed image and the entropy of the image have 1338/1256/1270 and 1156 distinct values, respectively. Consequently, the optical encoding of the input data has more than thousand levels resulting in a quasi-continuous amplitude distribution. The output is obtained with a single linear classifier using a percentage ratio between the training and test set of 70/30 where each banknote is randomly used five times. For each wavelength, the performance of the network is calculated using 100 different resamples of the test and training sets obtained by Monte Carlo cross-validation. In this task, the results obtained both with the 18 physical nodes alone and with the 18 physical nodes plus 126 virtual nodes are studied. The 126 virtual nodes are obtained by applying seven additional functions (median, standard deviation, skewness, kurtosis, geometric and harmonic mean) to the 1296 pixels contained in the image rectangle of each grating coupler. Applying the linear classifier to the raw data, we obtain a theoretical limit of the classification rate of $(97.57 \pm 0.07)\%$. Considering only the 18 physical nodes, the PELM does not significantly improve the classification rate. In particular, a maximum value of $(97.92 \pm 0.05)\%$ is reached for a wavelength close to the blue limit of the resonance region, i.e., 1560.9 nm. The use of virtual nodes greatly improves the performance of the PELM for wavelengths in-resonance and out-of-resonance. In particular, the classification rate reaches a maximum value of $(99.00 \pm 0.04)\%$ in resonance at 1561.3 nm. For comparison, Lupo *et al.*[19] achieve a classification rate of 99.4% and 99.7% with a photonic- and software-implemented ELM, respectively.

## IV. CONCLUSION

In this work, we experimentally demonstrated an Extreme Learning Machine architecture that exploits light scattered by an array of pair-coupled microresonators. The information is encoded in the PIC by modulating the amplitude of the input signals through Mach–Zehnder interferometers actuated by microheaters. The hidden layer of the network requires no training of its response. The randomness of the connections among the different microresonators in the array is ensured by stochastic fabrication errors. The readout layer is implemented digitally by using a linear classifier, i.e., a regularized linear regression. In the training procedure, it determines the optimal weights that allow replicating the target. We experimentally demonstrated that the PELM solves the logical AND, OR, and XOR operations, showing an increase in performance for in-resonance input wavelength. For the XOR operation, we studied the performance as a function of the number of physical nodes, considering all the different spatial configurations. In the best scenario, a combination of the outputs of three microresonators is sufficient to match the performance of the 18 microresonator array. Furthermore, it was shown that in special cases the output of a single microresonator of the array can solve the XOR without using the linear classifier in the readout layer.

In addition, we experimentally demonstrated that the network can handle multi-level tasks such as iris flower classification and banknote authentication. In the case of flowers, considering only the 18 physical nodes vs the input information, the maximum classification rate is $(97.2 \pm 0.2)\%$ vs $(96.6 \pm 0.3)\%$ with a single classifier and $(96.8 \pm 0.2)\%$ vs $(83.6 \pm 0.5)\%$ with a classifier for each species in the winner takes all the scheme. The study of the performance as a function of the number of physical nodes shows that in the best configuration, the outputs of three microresonators are sufficient to achieve the classification rate of 18 microresonators. However, the average of all spatial configurations shows an improvement in performance over the raw data only with at least nine microresonators. In the case of banknotes, the maximum classification rate with 18 physical nodes is $(97.92 \pm 0.05)\%$ vs $(97.57 \pm 0.07)\%$. Otherwise, considering both 18 physical nodes and 126 virtual nodes, the maximum classification rate increases to $(99.00 \pm 0.04)\%$ vs $(97.57 \pm 0.07)\%$.

Our PELM suffers from two main limitations. The first is the low operational speed due to both the image acquisition and the input data encoding. In fact, the use of an infrared camera requires a steady state in the microresonator array to avoid thermal cross-talk. As a result, an average time of 500 ms is required between two different inputs. The second is the software implementation of the ridge regression readout. This needs the storage of the output images, which rapidly gets very large. Furthermore, processing a large set of images requires a long computational time, especially when both physical and virtual nodes are used. Both limitations can be overcome by considering the integration of the readout layer in the PIC. For example, coupling fast photodetectors to the microresonators would permit an electronic reading of the array status. In addition, a linear classifier can be integrated in the PIC by connecting the output of the microrings to a cascade of MZIs followed by PSs,[36] thus obtaining an all-chip PELM. Another possible improvement of the PELM is a different encoding procedure. Since a resonant system is particularly sensitive not only to the intensity of the optical signal but also to its phase. An improvement in the performance of the PELM can be achieved by considering a complex encoding[2] where both the phase and amplitude of the optical field are used.

## ACKNOWLEDGMENTS

We gratefully thank Professor Claudio Conti and Dr. Davide Pierangeli of the Sapienza University of Rome for useful suggestions and interesting discussions. We also thank Dr. Alessio Lugnan and Professor Paolo Bettotti for their useful insights and valuable

14 September 2023 07:13:43

comments. S. Biasi acknowledges the co-financing of the European Union FSE-REACT-EU, PON Research and Innovation 2014–2020 DM1062/2021. R. Franchi acknowledges the co-financing of PAT through the Q@TN joint lab. This project has received funding from the European Research Council (ERC) under the European Union's Horizon 2020 research and innovation program (Grant Agreement No. 788793, BACKUP) and from the MIUR under the project PRIN PELM (Grant No. 20177 PSCKT).

## AUTHOR DECLARATIONS

### Conflict of Interest

The authors have no conflicts to disclose.

### Author Contributions

S.B. and R.F. contributed equally to this work.

**Stefano Biasi**: Conceptualization (lead); Data curation (equal); Formal analysis (equal); Investigation (equal); Methodology (equal); Software (equal); Validation (equal); Writing – original draft (lead). **Riccardo Franchi**: Conceptualization (equal); Data curation (equal); Formal analysis (equal); Investigation (equal); Methodology (equal); Software (equal); Validation (equal); Writing – review & editing (equal). **Lorenzo Cerini**: Data curation (equal); Formal analysis (supporting); Investigation (supporting); Software (equal); Writing – review & editing (supporting). **Lorenzo Pavesi**: Conceptualization (supporting); Funding acquisition (lead); Methodology (equal); Project administration (lead); Supervision (lead); Validation (equal); Writing – review & editing (lead).

## DATA AVAILABILITY

The data that support the findings of this study are available from the corresponding author upon reasonable request.

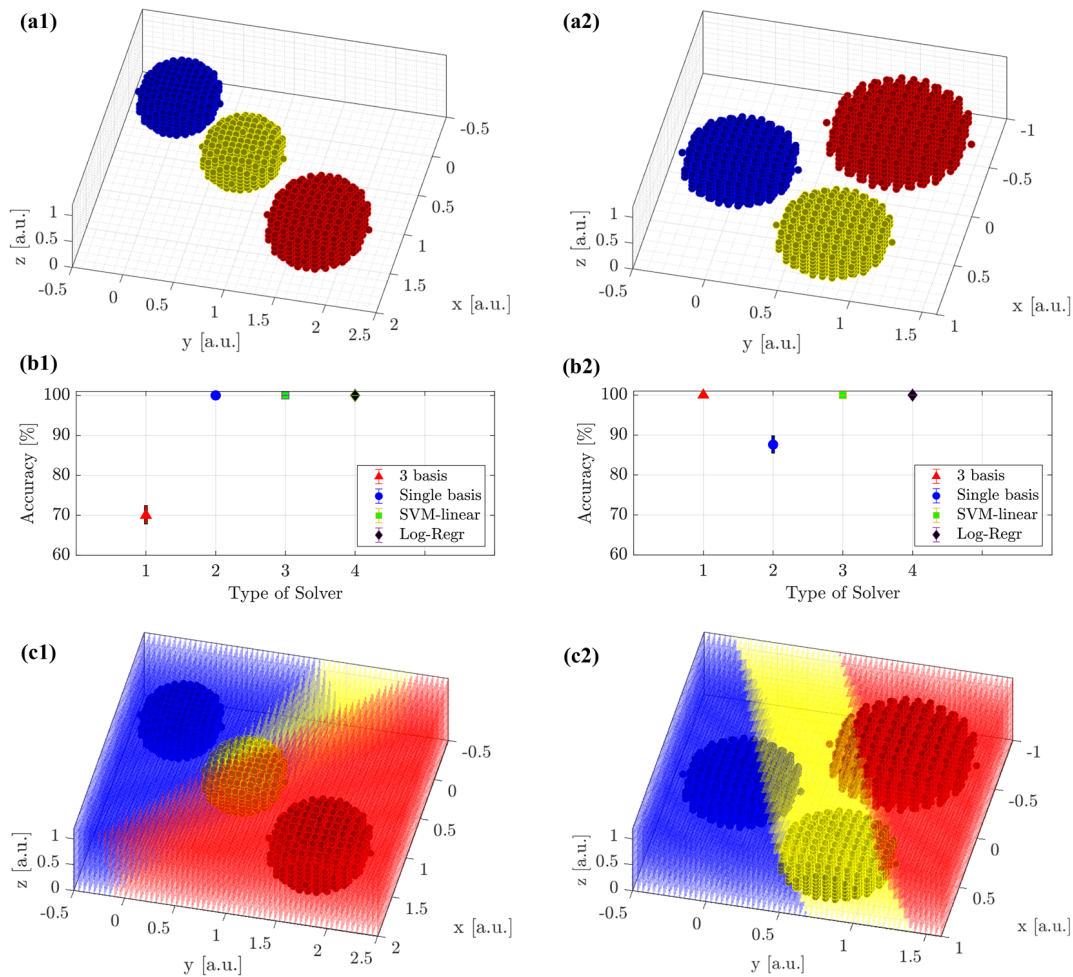

**FIG. 5.** Graphs (a1) and (b1) show two linear configurations of the database, which consists of three spheres made up of 925 triplets of real numbers representing their spatial coordinates. The three spheres are classified by three colors: blue, yellow, and red. The accuracy as a function of the solver type for the two configurations (a1) and (a2) are shown in plots (b1) and (b2), respectively. The three-dimensional plots (c1) and (c2) show the output result obtained by applying the solver that gives the worst accuracy in (b1) and (b2) to a generic matrix of three-dimensional points.

## APPENDIX A: LINEAR CLASSIFIERS

In this section, we compare the results of a single classifier with three classifiers in the winner takes all scheme. To visualize the results, we use a simple database consisting of three spheres. They consist of a set of 925 points in the 3D-plane and they are represented in Fig. 5 with three different colors defining their species: blue, yellow and red. As a result, each point in the database is a triple of real values that provide the spatial coordinates. We study two different spatial configurations of the three linearly separable spheres shown in Figs. 5(a1) and 5(a2), respectively. The dataset is divided into training and test with a ratio of 72/28. Training and testing are performed on the raw data using digital hardware. The multiclass classification is calculated in terms of accuracy by the single linear classifier, the three linear classifiers, a linear support vector machine (SVM), and a multinomial logistic regression software. We used Matlab[57] for all the implementations. Specifically, the linear classifiers were obtained by exploiting the *ridge* function with a regularization parameter $\lambda$ defined by a five-fold cross-validation using the *fitlinear* function. In contrast, SVM-linear and multinomial logistic regression were implemented using the *fitcecoc* function by defining the learner with the corresponding *templateLinear* function. For a single classifier (single basis) we use the target permutation which gives the best accuracy.

The accuracy as a function of the type of solver is shown in Figs. 5(b1) and 5(b2) for the spatial configuration represented in (a1) and (a2), respectively. Here, the red upward-triangles, blue dots, green squares, and black diamonds are the results obtained with the three linear classifiers (3 basis), single classifier (Single basis), SVM-linear, and multinomial logistic regression (Log-Regr), respectively. In the spatial configuration of three spheres in a row, shown in Fig. 5(a1), the three linear classifiers in the winner takes all scheme configuration fail to classify the three species by showing a value of 70.2%. In contrast, the other three solvers solve the task with an accuracy of 100%. Note that the single classifier works. In contrast, in the configuration of Fig. 5(a2), the single linear classifier fails to classify the three species perfectly, showing a value of 87.7%. While the three linear classifiers solve the task perfectly as the SVM linear and multinomial logistic regression. In Fig. 5, the graphs (c1) and (c2) show for a three-dimensional point matrix the classification result obtained in the worst accuracy case for configuration (a1) and (a2), respectively. In (c1) the three linear classifiers do not perfectly distinguish the yellow sphere by assigning red and blue points to it. Differently, in (c2), the single classifier does not perfectly distinguish the red and blue spheres by assigning yellow dots to them. Thus, the use of a ridge-based single classifier or multiple ridge-based classifiers in the winner takes all scheme can lead to completely different accuracy results and is highly dependent on the database under test.

## APPENDIX B: DESIGN PARAMETERS OF THE PELM ARCHITECTURE

Table II shows the nominal parameters used to design the microresonator array. It shows the values of the racetrack microresonator, which is composed of four Euler curves and two straight sections, as well as the values of the gaps. In particular, the gaps of the microresonator-bus waveguide (M-W), microresonator-microresonator (M-M), microresonator-bus waveguide-microresonator (M-W-M) and microresonator-tap (M-T) connected to the grating coupler are listed. As mentioned in the main text, the reported values were used for each of the three microresonator columns in the CROW configuration [see Fig. 1(b)]. The coupling efficiency (i.e., the square of the coupling coefficient) corresponding to the gap values are 0.10, 0.099, 0.10 and 0.005 for M-W, M-M, M-W-M and M-T, respectively. The microresonators thus designed have a field enhancement of about 9. Propagation and insertion losses were estimated by measuring the transmission of straight waveguides of different nominal lengths. The values obtained for an input wavelength of 1550 nm are 2 dB/cm for propagation losses and 6 dB for insertion losses (i.e., about 3 dB for each grating coupler).

## APPENDIX C: MULTINOMIAL LOGISTIC REGRESSION ON THE IRIS FLOWER DATABASE

In this section, we discuss the results of the Iris flower classification computed using the multinomial logistic regression in the readout layer. As in the case of the other linear classifiers, the distribution of both the raw and output data between the training and test sets is 70% and 30%, respectively.

Figures 6(a) and 6(b) show the classification rate as a function of incident wavelength obtained by using 18 and 5 physical nodes as output, respectively. Precisely, the blue and red triangles show the mean values over the 100 repetitions with the corresponding variances indicated by the error bars for the single basis linear classifier (ridge regression) and the multinomial logistic regression, respectively. The blue and red horizontal rectangles show our theoretical limits, i.e., the mean and variance of the classification rate performed on the raw input data.

The single basis provides higher classification rates than the multinomial logistic regression for both the 18 and the five physical nodes. In particular, with a percentage ratio between the training and test set of 70/30, this occurs even at the theoretical limit of the raw data: $(96.6 \pm 0.3)\%$ vs $(95.5 \pm 0.4)\%$. In the case of 18 physical nodes, the readout of the output through multinomial logistic regression does not provide a net performance improvement, showing a maximum value close to the theoretical limit of $(95.9 \pm 0.3)\%$ out-of-resonance at 1560.9 nm. This lack of improvement is due to overfitting during training procedure. In fact, by reducing the number of nodes to 5, the multinomial logistic regression shows a significant improvement in the output response of the network. Specifically, the data follow a trend similar to that of the single basis, with a maximum value of $(96.3 \pm 0.2)\%$ at 1561.5 nm. This value does not represent the absolute maximum that can be achieved with five physical nodes. In fact, the five rings are chosen randomly without studying the permutation on the 18 microresonators that provides the best performance. Note that in the case of three linear classifiers [Fig. 4(b)], the output obtained by ridge regression shows a significant improvement in performance, even with 18 physical nodes. In this case, the network does not suffer from overfitting during training since parameters of the three classifiers are determined by considering only one class against the other two.

**TABLE II.** PELM network design parameters.

| Racetrack microresonator | | | Coupling gaps | | | |
|---|---|---|---|---|---|---|
| Minimum Euler radius | Euler angle | Straight sections | M-W | M-M | M-W-M | M-T |
| 15 $\mu$m | $\frac{\pi}{2}$ rad | 4 $\mu$m | 289 nm | 260 nm | 287 nm | 412 nm |

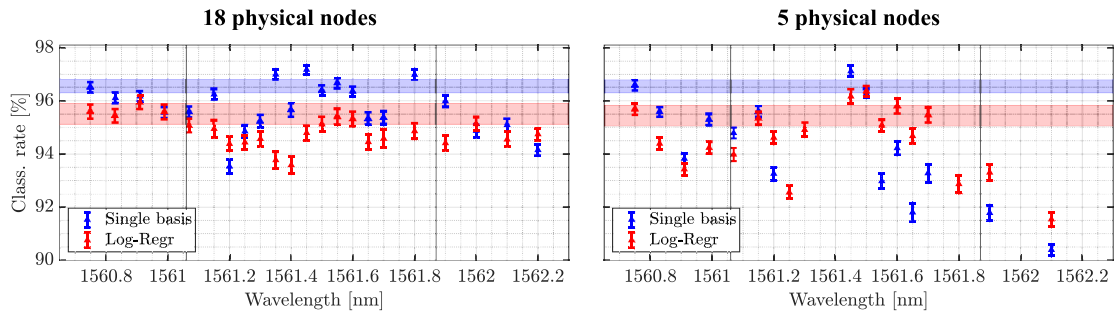

**FIG. 6.** Classification rate as a function of input wavelength, computed using a single linear classifier (ridge regression) and multinomial logistic regression for 18 (a) and 5 (b) physical nodes, respectively. The blue and red horizontal rectangle highlight the theoretical limit obtained by applying the readout procedure to the raw input data. With 18 physical nodes, the multinomial logistic regression training suffers from overfitting and shows no significant improvement in output over input. By reducing the number of physical nodes to 5, the logistic regression output improves network performance by showing an in-resonance maximum.

## REFERENCES

[1] N. L. Kazanskiy, M. A. Butt, and S. N. Khonina, "Optical computing: Status and perspectives," Nanomaterials **12**(13), 2171 (2022).

[2] H. Zhang, M. Gu, X. D. Jiang, J. Thompson, H. Cai, S. Paesani, R. Santagati, A. Laing, Y. Zhang, M. H. Yung, Y. Z. Shi, F. K. Muhammad, G. Q. Lo, X. S. Luo, B. Dong, D. L. Kwong, L. C. Kwek, and A. Q. Liu, "An optical neural chip for implementing complex-valued neural network," Nat. Commun. **12**(1), 457 (2021).

[3] Y. Bai, X. Xu, M. Tan, Y. Sun, Y. Li, J. Wu, R. Morandotti, A. Mitchell, K. Xu, and D. J. Moss, "Photonic multiplexing techniques for neuromorphic computing," Nanophotonics **12**(5), 795–817 (2023).

[4] F. Ashtiani, A. J. Geers, and F. Aflatouni, "An on-chip photonic deep neural network for image classification," Nature **606**(7914), 501–506 (2022).

[5] X. Xu, M. Tan, B. Corcoran, J. Wu, A. Boes, T. G. Nguyen, S. T. Chu, B. E. Little, D. G. Hicks, R. Morandotti, A. Mitchell, and D. J. Moss, "11 TOPS photonic convolutional accelerator for optical neural networks," Nature **589**(7840), 44–51 (2021).

[6] H. H. Zhu, J. Zou, H. Zhang, Y. Z. Shi, S. B. Luo, N. Wang, H. Cai, L. X. Wan, B. Wang, X. D. Jiang, J. Thompson, X. S. Luo, X. H. Zhou, L. M. Xiao, W. Huang, L. Patrick, M. Gu, L. C. Kwek, and A. Q. Liu, "Space-efficient optical computing with an integrated chip diffractive neural network," Nat. Commun. **13**(1), 1044 (2022).

[7] P. Xu and Z. Zhou, "Silicon-based optoelectronics for general-purpose matrix computation: A review," Adv. Photonics **4**(04), 044001 (2022).

[8] B. J. Shastri, A. N. Tait, T. Ferreira de Lima, W. H. P. Pernice, H. Bhaskaran, C. D. Wright, and P. R. Prucnal, "Photonics for artificial intelligence and neuromorphic computing," Nat. Photonics **15**(2), 102–114 (2021).

[9] K. Liao, T. Dai, Q. Yan, X. Hu, and Q. Gong, "Integrated photonic neural networks: Opportunities and challenges," ACS Photonics **10**, 2001–2010 (2023).

[10] Y. Shen, N. C. Harris, S. Skirlo, M. Prabhu, T. Baehr-Jones, M. Hochberg, X. Sun, S. Zhao, H. Larochelle, D. Englund, and M. Soljačić, "Deep learning with coherent nanophotonic circuits," Nat. Photonics **11**(7), 441–446 (2017).

[11] D. E. Rumelhart, G. E. Hinton, and R. J. Williams, "Learning representations by back-propagating errors," Nature **323**(6088), 533–536 (1986).

[12] P. van der Smagt and G. Hirzinger, "Why feed-forward networks are in a bad shape," in *ICANN 98: Proceedings of the 8th International Conference on Artificial Neural Networks, Skövde, Sweden, 2–4 September 1998* (Springer, 1998), pp. 159–164.

[13] M. Lukoševičius and H. Jaeger, "Reservoir computing approaches to recurrent neural network training," Comput. Sci. Rev. **3**(3), 127–149 (2009).

[14] C. Gallicchio, A. Micheli, and L. Pedrelli, "Deep reservoir computing: A critical experimental analysis," Neurocomputing **268**, 87–99 (2017).

[15] H. Jaeger, "The 'echo state' approach to analysing and training recurrent neural networks-with an erratum note," GMD Technical Report, German National Research Center for Information Technology, Bonn, Germany, 2001, Vol. 148, pp. 13–34.

[16] G.-B. Huang, Q.-Y. Zhu, and C.-K. Siew, "Extreme learning machine: Theory and applications," Neurocomputing **70**(1–3), 489–501 (2006).

[17] G.-B. Huang, H. Zhou, X. Ding, and R. Zhang, "Extreme learning machine for regression and multiclass classification," IEEE Trans. Syst., Man, Cybern. Part B **42**(2), 513–529 (2011).

[18] D. Pierangeli, G. Marcucci, C. Conti *et al.*, "Photonic extreme learning machine by free-space optical propagation," Photonics Res. **9**(8), 1446–1454 (2021).

[19] A. Lupo, L. Butschek, and S. Massar, "Photonic extreme learning machine based on frequency multiplexing," Opt. Express **29**(18), 28257–28276 (2021).

[20] A. Saade, F. Caltagirone, I. Carron, L. Daudet, A. Drémeau, S. Gigan, and F. Krzakala, "Random projections through multiple optical scattering: Approximating kernels at the speed of light," in *2016 IEEE International Conference on Acoustics, Speech and Signal Processing (ICASSP)* (IEEE, 2016), pp. 6215–6219.

[21] S. Sunada, K. Kanno, and A. Uchida, "Using multidimensional speckle dynamics for high-speed, large-scale, parallel photonic computing," Opt. Express **28**(21), 30349–30361 (2020).

[22] S. Ortín, M. C. Soriano, L. Pesquera, D. Brunner, D. San-Martín, I. Fischer, C. R. Mirasso, and J. M. Gutiérrez, "A unified framework for reservoir computing and extreme learning machines based on a single time-delayed neuron," Sci. Rep. **5**(1), 14945 (2015).

[23] K. Vandoorne, P. Mechet, T. Van Vaerenbergh, M. Fiers, G. Morthier, D. Verstraeten, B. Schrauwen, J. Dambre, and P. Bienstman, "Experimental demonstration of reservoir computing on a silicon photonics chip," Nat. Commun. **5**, 3541 (2014).

[24]M. Nakajima, K. Tanaka, and T. Hashimoto, "Scalable reservoir computing on coherent linear photonic processor," Commun. Phys. **4**(1), 20 (2021).

[25]P. Antonik, N. Marsal, and D. Rontani, "Large-scale spatiotemporal photonic reservoir computer for image classification," IEEE J. Sel. Top. Quantum Electron. **26**(1), 7700812 (2020).

[26]M. Borghi, S. Biasi, and L. Pavesi, "Reservoir computing based on a silicon microring and time multiplexing for binary and analog operations," Sci. Rep. **11**(1), 15642 (2021).

[27]L. Jia, Q. Wu, X. Sui, Q. Chen, G. Gu, L. Wang, and S. Li, "Research progress in optical neural networks: Theory, applications and developments," PhotoniX **2**(1), 5 (2021).

[28]B. Stefano, R. Franchi, D. Bazzanella, and L. Pavesi, "On the effect of the thermal cross-talk in a photonic feed-forward neural network based on silicon microresonators," Front. Phys. **10**, 1350 (2022).

[29]M. Mancinelli, M. Borghi, P. Bettotti, J. M. Fedeli, and L. Pavesi, "An all optical method for fabrication error measurements in integrated photonic circuits," J. Lightwave Technol. **31**(14), 2340–2346 (2013).

[30]R. A. Fisher, "The use of multiple measurements in taxonomic problems," Ann. Eugen. **7**(2), 179–188 (1936).

[31]D. Dua and C. Graff, UCI machine learning repository, 2017.

[32]M. Kuhn, K. Johnson *et al.*, *Applied Predictive Modeling* (Springer, 2013), Vol. 26.

[33]J. Zurada, *Introduction to Artificial Neural Systems* (West Publishing Co., 1992).

[34]C. M. Bishop and N. M. Nasrabadi, *Pattern Recognition and Machine Learning* (Springer, 2006), Vol. 4.

[35]L. Larger, M. C. Soriano, D. Brunner, L. Appeltant, J. M. Gutierrez, L. Pesquera, C. R. Mirasso, and I. Fischer, "Photonic information processing beyond Turing: An optoelectronic implementation of reservoir computing," Opt. Express **20**(3), 3241–3249 (2012).

[36]H. Zhou, J. Dong, J. Cheng, W. Dong, C. Huang, Y. Shen, Q. Zhang, M. Gu, C. Qian, H. Chen, Z. Ruan, and X. Zhang, "Photonic matrix multiplication lights up photonic accelerator and beyond," Light: Sci. Appl. **11**(1), 30 (2022).

[37]The MathWorks Inc., Matlab version: 9.13.0 (r2022b), 2022.